# The electrocaloric effect of lead-free $Ba_{1-y}Ca_yTi_{1-x}Hf_xO_3$ from direct and indirect measurements

David Gracia[1,2,a], Sara Lafuerza[1,2,a], Javier Blasco[1,2,a], and Marco Evangelisti[1,2,a]

[1] Instituto de Nanociencia y Materiales de Aragón (INMA), CSIC-Universidad de Zaragoza, 50009 Zaragoza, Spain

[2] Departamento de Física de la Materia Condensada, CSIC-Universidad de Zaragoza, 50009 Zaragoza, Spain

[a] Authors to whom correspondence should be addressed: davidg@unizar.es, lafuerza@unizar.es, jbc@unizar.es and evange@unizar.es

**ABSTRACT**

We report on the dielectric and electrocaloric properties of $Ba_{1-y}Ca_yTi_{1-x}Hf_xO_3$ for compositions $0.12 < x < 0.18$ and $y = 0.06$, as well as $x = 0.15$ and $0 < y < 0.15$, synthesized by the conventional solid-state reaction method. The addition of Hf/Ca broadens the ferroelectric-paraelectric phase transition, while moving it toward room temperature. Two interferroelectric transitions are seen to converge, together with the ferroelectric-paraelectric phase transition, at ca. 335 K for $0.12 < x_c < 0.135$ and $y = 0.06$. Consistently with the dielectric properties, the electrocaloric effect maximizes closer to room temperature with increasing Hf/Ca substitutions, which promote larger temperature spans. The electrocaloric responsivity gradually decreases from 0.2 to 0.1 K mm kV$^{-1}$ with the addition of Hf/Ca. A homemade quasi-adiabatic calorimeter is employed to measure "directly" the electrocaloric data, which are also calculated from polarization-versus-electric-field cycles using "indirect" standard procedures. The comparison between measured and calculated values highlights the importance of having access to direct methods for a reliable determination of the electrocaloric effect.

# I. INTRODUCTION

There is a great interest in the search for refrigeration systems that can replace the current technologies based on the vapor compression of environmentally harmful gases. Solid-state refrigeration based on caloric materials is a promising alternative because it is free of greenhouse gases, can achieve higher energy efficiencies than current systems and can be integrated in smaller devices.[1] The electrocaloric effect (ECE), which consists in an adiabatic and reversible thermal change following the application or removal of an electric field, is experiencing a renewed interest since the discovery of a giant ECE in thin films of $PbTi_{0.05}Zr_{0.95}O_3$.[2] The effect is usually stronger in the vicinity of a ferroelectric (FE)-paraelectric phase transition, that is, near the Curie temperature, $T_c$.[3] Therefore, ferroelectrics with $T_c$ close to room temperature are of special interest for domestic refrigeration. However, most commercial ferroelectrics show transitions at too high temperatures for this purpose. The tuning of $T_c$ toward room temperature can be attained by chemical doping, as successfully reported for lead-based perovskites that have shown a high ECE at temperatures close to room temperature.[4] However, lead is a harmful element that should be avoided for industrial applications, therefore FE ceramics such as $BaTiO_3$ (BTO) and its derivatives should be preferred.[5]

BTO is a well-known FE with $T_c$ = 393 K.[6] The FE-paraelectric phase transition is coupled to a structural transition, viz., from the paraelectric cubic phase (space group *Pm-3m*) to the FE tetragonal one (*P4mm*). BTO undergoes two more structural transitions on cooling. Between $T_O$ = 273 K and $T_R$ = 183 K, BTO is orthorhombic (*Amm2*). Below $T_R$, it adopts a rhombohedral structure (*R3m*). Each of these two transitions is accompanied by the reorientation of the polarization direction. Previous works on BTO reported ECE maxima near $T_c$ and $T_O$,[7,8] while first principles calculations predicted the ECE to spike at all three FE transitions.[9] The isovalent substitution of Ti sites with a tetravalent cation $M^{4+}$, such as $Zr^{4+}$, decreases $T_c$, while $T_O$ and $T_R$ are increased.[10] At critical compositions, $x_c$, the phase transitions converge within a region of the phase diagram where multiple phases coexist in thermodynamic equilibrium.[10] Critical compositions were studied because they exhibit high piezoelectric coefficients and large low-field unipolar strains.[11]

Above $x_c$, doping usually leads to the formation of relaxor compounds, where polar nanoregions are thought to form instead of long-range FE ordering.[12] With respect to conventional FE materials, relaxors can widen the useful temperature range,[13] and reduce electrical and thermal losses.[14] For this reason, the ECE in $BaTi_{1-x}M_xO_3$ has been recently studied for M = $Zr^{4+}$, $Hf^{4+}$ and $Sn^{4+}$.[15-17] Besides, Ca doping for Ba in $BaTi_{1-x}M_xO_3$ also affects the dielectric properties.[18] Despite a few works focused on the ECE of Zr- and Ca-doped compositions,[13,19,20] no systematic studies of the ECE dependance on the simultaneous replacement of Ti by Hf and Ba by Ca have been reported.

We report the $BaTiO_3$-$CaTiO_3$-$BaHfO_3$ pseudo-phase diagram and its dielectric and electrocaloric properties in the proximity of its phase convergence region. This system has been chosen because FE transitions with $T_c$ = 313 and 293 K have been reported for the compositions $Ba_{0.94}Ca_{0.06}Ti_{0.8}Hf_{0.2}O_3$ and $Ba_{0.85}Ca_{0.15}Ti_{0.85}Hf_{0.15}O_3$, respectively.[11,18] Depending on the composition, the materials undergo consecutive transitions within a narrow temperature span, ultimately leading to the formation of a phase convergence region at $x_c$. To optimize the electrocaloric properties, we focus on two $Ba_{1-y}Ca_yTi_{1-x}Hf_xO_3$ series, fixing either $x$- or $y$-values to account for the effects of the different substitutions. We combine direct ECE measurements, using a homemade quasi-adiabatic calorimeter, together with indirect ECE measurements, derived from polarization-versus-electric-field cycles. Our results show that the agreement between both methods is qualitative but not quantitative because of the limitations inherent in the indirect estimates.

## II. EXPERIMENTAL

Two sets of $Ba_{1-y}Ca_yTi_{1-x}Hf_xO_3$ (BCTH-$y/x$) samples were prepared by solid-state reaction. In the first set, $y$ was fixed to 0.06 and $x$ had the values of 0.12, 0.135, 0.15, 0.165 and 0.18 (BCTH-6/$x$). In the second series, $x$ was fixed to 0.15 and $y$ took the values of 0, 0.06, 0.105 and 0.15 (BCTH-$y$/15). Therefore, the composition $Ba_{0.94}Ca_{0.06}Ti_{0.85}Hf_{0.15}O_3$ (BCTH-6/15) belongs to both series. Stoichiometric amounts of $BaCO_3$, $CaCO_3$, $TiO_2$ and $HfO_2$ were mixed, ground, and heated at 1280º C for 3 h. The resulting powder was reground, pressed into disc-shaped pellets (diameter of ∼ 8 mm and thickness ∼ 1.2 mm), and sintered at 1450º C for 3 h in air.

The samples were characterized by powder X-ray diffraction (XRD) at room temperature with a Rigaku D-Max system using Cu K$\alpha_{1,2}$ wavelengths. The X-ray diffraction patterns were analyzed by the Rietveld method using the Fullprof program.[21]

The relative dielectric permittivity was measured on disc-shaped pellets as a function of temperature between 300 and 420 K in a tubular furnace, employing a homemade coaxial line inset. Silver paint was applied to the disc surface for proper electrical contact. The complex dielectric permittivity of the samples was measured using an impedance analyzer (WayneKerr Electronics 6500B), applying excitations of 1 V in a frequency range between 100 Hz and 4 MHz. The data were taken on cooling at −1 K min$^{-1}$.

Electrical polarization versus electric-field loops measurements were performed on similar disc-shaped pellets using silver paint in sandwich geometry. The sample polarization was recorded using a commercial polarization analyzer (aixACCT Systems Easy Check 300) at 10 Hz, using sinusoidal excitation voltages and electric-field amplitudes of 20 − 25 kV cm$^{-1}$ in a silicone oil bath. A homemade sample holder equipped with a heater resistance was used in the temperature range from room temperature up to 420 K. The polarization data were employed for calculating the ECE, following known procedures based on Maxwell's relations.

The direct determination of the ECE requires instruments that are not readily available, being only a few groups that have developed custom-made setups.[1,22] Modified differential scanning calorimetry is the most used technique for direct ECE measurements.[23] Herein, we make use of a homemade quasi-adiabatic calorimeter, following a similar approach as per other groups.[24] The temperature was stabilized in 1 K-steps, from room temperature to 365 K. Disc-shaped and silver-paste electrodes were used in sandwich geometry. Voltages up to $V = 2$ kV were applied, the electrical field acting on the sample being $E = V/d$, where $d$ is the sample thickness. During the whole measurement, the sample temperature was monitored with a K-type thermocouple, located on the grounded face of the sample. Further description is provided in the Appendix together with example raw measurements.

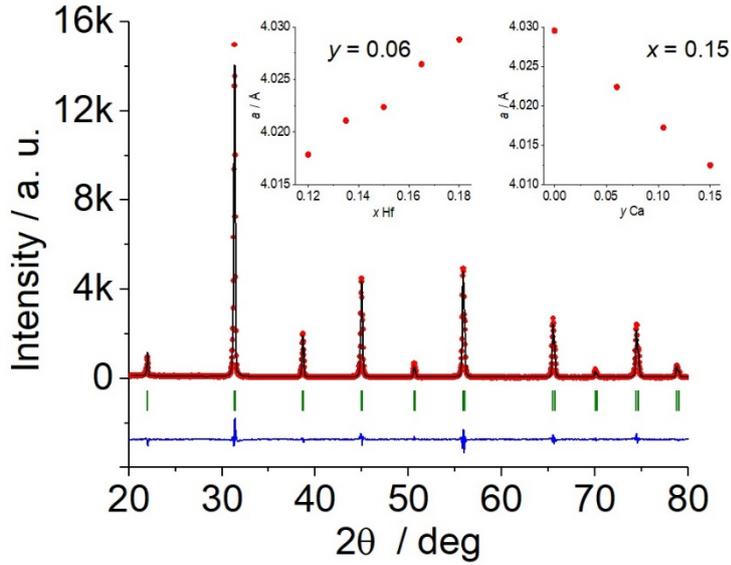

**FIG. 1**. Room-temperature XRD for BCTH-6/15, together with its Rietveld refinement (*Pm-3m* space group). The insets show the cubic cell parameter a as a function of the Hf and Ca contents, respectively, for both series of compositions.

## III. RESULTS AND DISCUSSION

### A. Crystal structure at room temperature

The XRD patterns of all $Ba_{1-y}Ca_yTi_{1-x}Hf_xO_3$ ceramics studied in this work exhibit a perovskite single phase (see Fig. S1 for full data). Fig. 1 shows a representative XRD pattern with the Rietveld refinement for the intermediate composition BCTH-6/15. No peak splitting ascribed to tetragonal, orthorhombic, or rhombohedral phases is noticeable for any sample. Therefore, the XRD patterns can be refined using the cubic *Pm-3m* space group with very good reliability factors. The refined lattice parameter *a* increases linearly with the Hf content for the B-site substitution, while it decreases linearly with the Ca content for the A-site substitution. The linear behavior is attributed to the different radii of the cations present in this system, namely $Ti^{4+}$(0.605 Å), $Hf^{4+}$(0.71 Å), $Ba^{2+}$(1.61 Å) and $Ca^{2+}$(1.34 Å).[25] These features indicate that perfect solid solutions (Ti/Hf and Ba/Ca) are formed, and that a pseudo-cubic structure is observed in average since a true cubic symmetry would be inconsistent with the ferroelectric properties, described next.

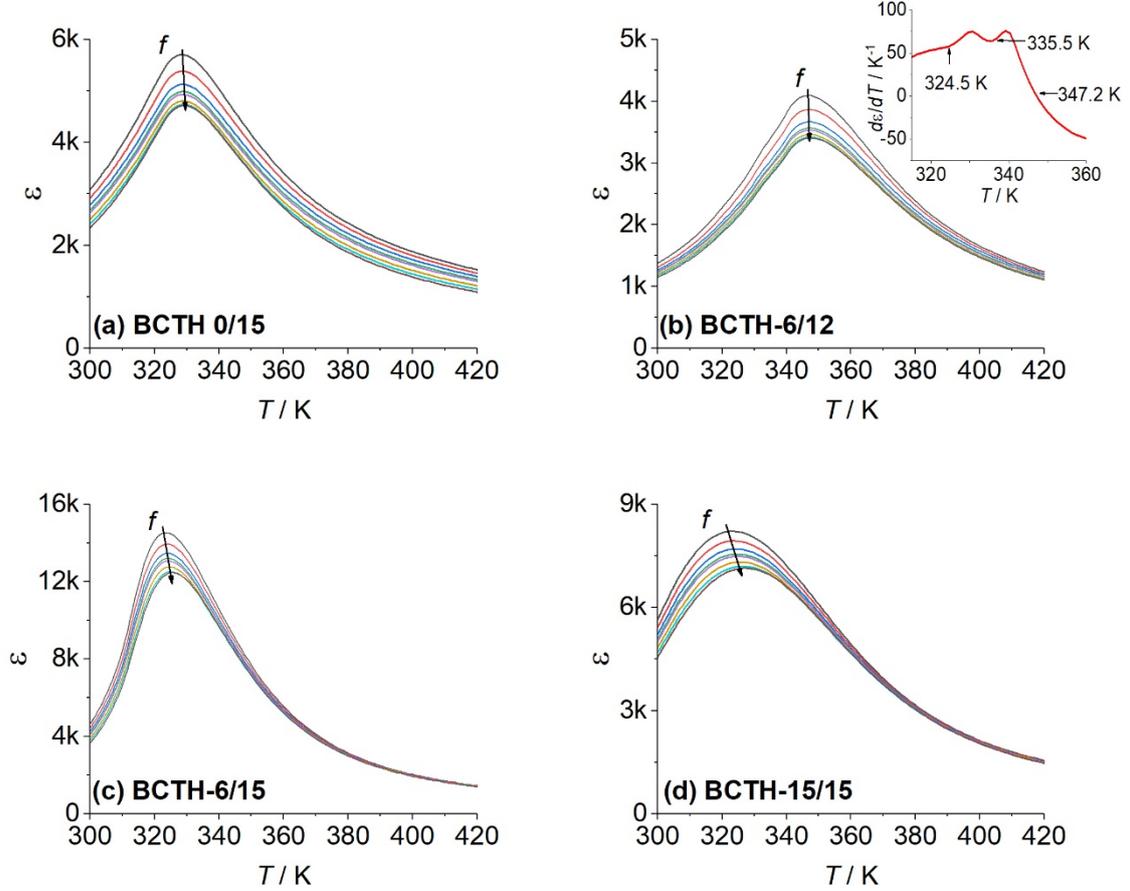

**FIG. 2**. Temperature and frequency dependances of the real component of the dielectric permittivity, measured between 100 Hz and 4 MHz, for four selected $Ba_{1-y}Ca_yTi_{1-x}Hf_xO_3$ compositions: BCTH-0/15 (a), BCTH-6/12 (b), BCTH-6/15 (c) and BCTH-15/15 (d). The inset shows the first derivative of the dielectric permittivity for BCTH-6/12, at 50 kHz.

**B. Dielectric properties**

The ferroelectric ordering of $Ba_{1-y}Ca_yTi_{1-x}Hf_xO_3$ can be studied by analyzing the real component of the dielectric permittivity $\varepsilon$, that we collect for temperatures ranging from room temperature to 420 K for both sets of compositions (Figs. 2, S2 and S3). As the Hf or Ca content increases, the temperature $T_m$ of the main anomaly moves toward room temperature and broadens, this behavior being characteristic of ferroelectrics with a diffuse phase transition.[12] The structural disorder can lead to electric field fluctuations and the formation of polar nanoregions, ultimately resulting in relaxor behavior that can be investigated by analyzing the frequency shift of the principal anomaly in $\varepsilon$ (Figs. 2, S2 and S3). We denote $\Delta T_m$, the difference between the temperatures of the maxima measured for the lowest (100 Hz) and highest (4 MHz) frequencies. As summarized in Table I for all compositions, $\Delta T_m$ tends to grow with increasing Hf/Ca substitutions, suggesting the onset of relaxor behavior. Other BTO ceramics show this feature for similar concentrations of isovalent substitutions, e.g., 0.25, 0.23 and 0.20 for the B-site substitutions $Zr^{4+}$, $Hf^{4+}$ or $Sn^{4+}$,[26-28] respectively.

**TABLE I**. For $Ba_{1-y}Ca_yTi_{1-x}Hf_xO_3$, columns denote the Hf or Ca content, the temperature $T_m$ for 50 kHz, the difference $\Delta T_m$ between $T_m$'s for the lowest (100 Hz) and highest (4MHz) frequencies, and Uchino and Nomura $\gamma$ exponent (see Eq. 2), respectively.

| BCTH-6/x | | | | BCTH-y/15 | | | |
|---|---|---|---|---|---|---|---|
| x Hf | $T_m$ (K) | $\Delta T_m$ (K) | $\gamma$ | y Ca | $T_m$ (K) | $\Delta T_m$ (K) | $\gamma$ |
| 0.12 | 347.2 | 1.0 | 1.47 | 0.00 | 328.9 | 0.5 | 1.30 |
| 0.135 | 338.9 | 0.8 | 1.62 | 0.06 | 324.5 | 2.0 | 1.70 |
| 0.15 | 324.5 | 2.0 | 1.70 | 0.105 | 324.8 | 2.4 | 1.76 |
| 0.165 | 317.7 | 1.8 | 1.71 | 0.15 | 324.8 | 3.9 | 1.87 |
| 0.18 | 306.2 | 1.9 | 1.75 | | | | |

For all compositions the transition peaks are much broader than in BTO, and the paraelectric region no longer obeys the Curie-Weiss dependance, i.e.,

$$\varepsilon = \frac{C}{T - \theta} \tag{1}$$

where $C$ is the Curie constant and $\theta$ is the Curie-Weiss temperature. To quantify the broadening, we fit the paraelectric region to the phenomenological expression by Uchino and Nomura,[29] i.e.,

$$\frac{1}{\varepsilon} - \frac{1}{\varepsilon_m} = \frac{(T - T_m)^\gamma}{C'} \tag{2}$$

where $\varepsilon_m$ corresponds to the maximum value of the dielectric permittivity, found at a temperature $T_m$, and $C'$ and $\gamma$ are fitting parameters. Specifically, $C'$ is a Curie-like constant and $\gamma$ gives information on the character of the transition. For $\gamma = 1$, the Curie-Weiss law for conventional ferroelectrics is recovered, while the limit $\gamma = 2$ indicates the so-called complete diffuse phase transition.[30] Typical values obtained for disordered systems are in the interval $1 < \gamma < 2$ and represent intermediate states.[12] The $\gamma$ exponents obtained for BCTH-y/x and shown in Table I are frequency independent (see fits in Figs. S2 and S3). It can be noted that $\gamma$ gradually increases from 1 to 2 with increasing Hf/Ca concentrations and hence structural disorder, which eventually lead to relaxor behavior.

The BCTH-6/12 stands apart from all other investigated compositions, being the only one showing two further anomalies below the main one (Fig. 2), which tentatively correspond to the interferroelectric transitions $T_O$ and $T_R$, respectively, as in BTO. Note that the Hf substitution increases both $T_O$ and $T_R$, while $T_c$ is reduced. The three temperatures eventually converge at ~ 335 K for $0.12 < x_c < 0.135$, leading to the phase convergence region for BCTH-6/$x$, in agreement with previous results.[11] Since each composition of BCTH-$y$/15 exhibits one maximum solely, the three transitions have likely already converged for BCTH-0/15, i.e., the first composition of this set. In this case, $T_c$ decreases slightly with Ca substitutions.

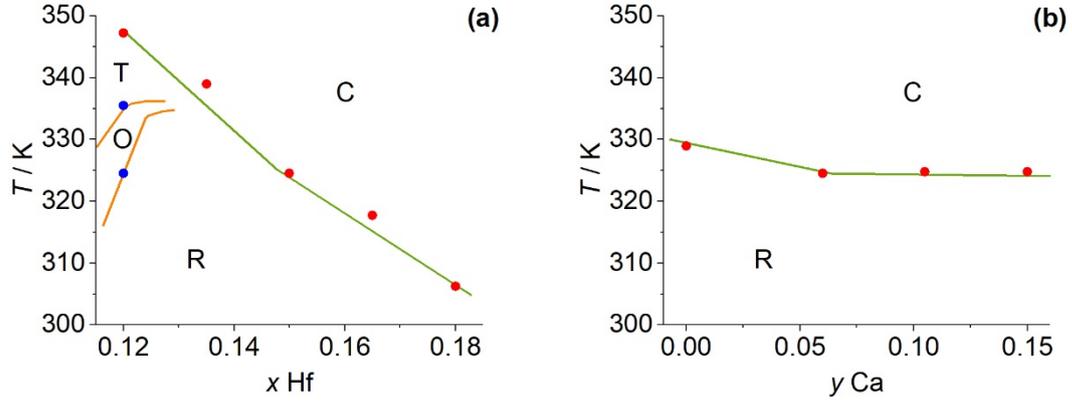

**FIG. 3**. Pseudo-phase diagrams for $Ba_{1-y}Ca_yTi_{1-x}Hf_xO_3$; (a) BCTH-6/$x$, i.e., fixed Ca and variable Hf content and (b) BCTH-$y$/15, i.e., fixed Hf and variable Ca content. Region labels: cubic (C), tetragonal (T), orthorhombic (O), rhombohedral (R).

Next, we determine the pseudo-phase diagrams for both sets, as a function of the Hf/Ca content, from the $T_m$ values measured at 50 kHz (Table I). Regions of the pseudo-phase diagrams are labeled in Fig. 3 according to the structures that BTO would exhibit correspondingly. One can observe that all compositions are located within the rhombohedral region at room temperature, although their average structure found by XRD is pseudo-cubic. This suggests that the distortions from the cubic *Pm-3m* space group in the XRD pattern are smaller than the instrumental resolution.

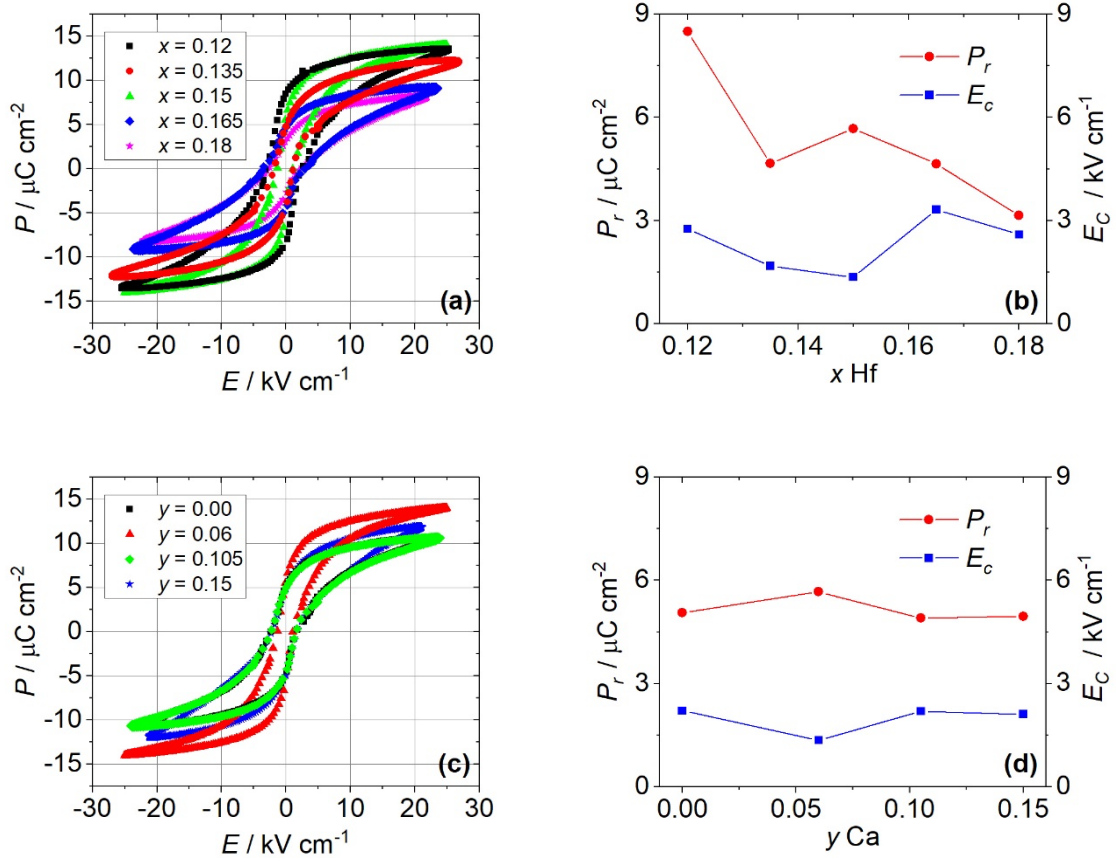

**FIG. 4**. Room-temperature versus electric-field loops and the dependance of room-temperature $P_r$ and $E_c$ on Ca/Hf concentrations, for BCTH-6/$x$ (a and b, respectively) and BCTH-$y$/15 (c and d respectively).

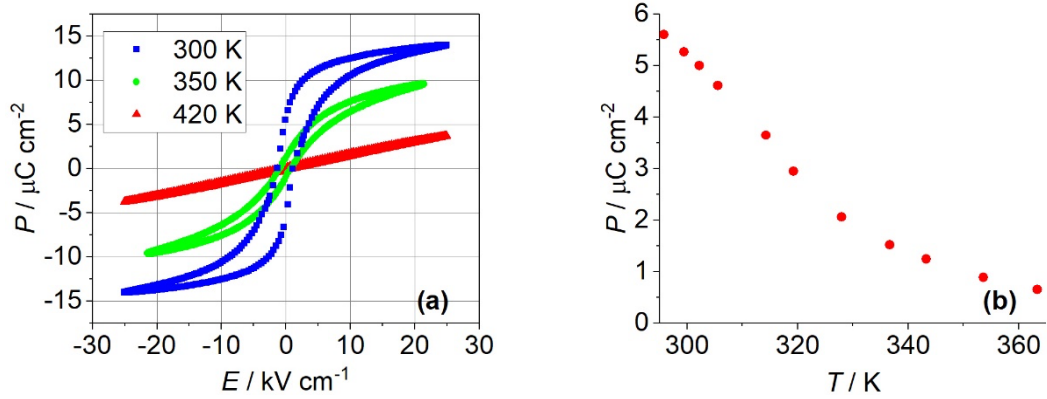

**FIG. 5**. For BCTH-6/15, polarization versus electric-field loops for selected temperatures (a) and temperature dependance of $P_r$ (b).

The composition dependance of the polarization versus electric-field loops and the evolution of the remanent polarization $P_r$ and the coercive field $E_c$, collected at room temperature, for both Hf and Ca substitutions, are shown in Fig. 4. For all compositions, the FE loops are softer than in BTO, as previously reported.[11,18] The increase of the Hf content tends to reduce the remanent polarization at room temperature, as expected since the FE transition temperature is decreased. The effect of the Ca substitution on the FE properties at room temperature is less pronounced. With increasing temperature, the polarization loops of BCTH-6/15 gradually fade, eventually disappearing above 400 K (Fig. 5). Similar results are obtained for all other compositions (Figs. S4 and S5).

**C. Electrocaloric properties**

For each investigated composition, direct measurements of the electrocaloric temperature change, $\Delta T$, for any field change $\Delta E = E - 0$, indicate that the maximum ECE occurs near the temperature, $T_m$, that also characterizes the maximum of the dielectric permittivity. As a representative example, BCTH-6/15 shows a maximum $\Delta T$ value of 0.2 K at 330 K, for $E = 16$ kV cm$^{-1}$ (Fig. 6). The temperature of the maximum $\Delta T$ also depends on the strength of the electric field, being higher as $E$ increases. For all other compositions, this shifting with $E$ tends to be more/less noticeable as the Hf or Ca content increases/decreases, as depicted in Fig. 7 (see also Figs. S6 and S7 for full data). This result can be ascribed to the fact that the transition temperature $T_c$ increases with the electric field $E$, as it is predicted for FE crystals.[31] Evidences of the shifting of the transition temperature in the ECE response have also been reported for BTO,[7,8] and relaxor ferroelectrics, namely (Na, Bi, K, Sr)(Ti, Nb)O$_3$ and Na$_{0.5}$Bi$_{0.5}$TiO$_3$ ceramics.[32,33]

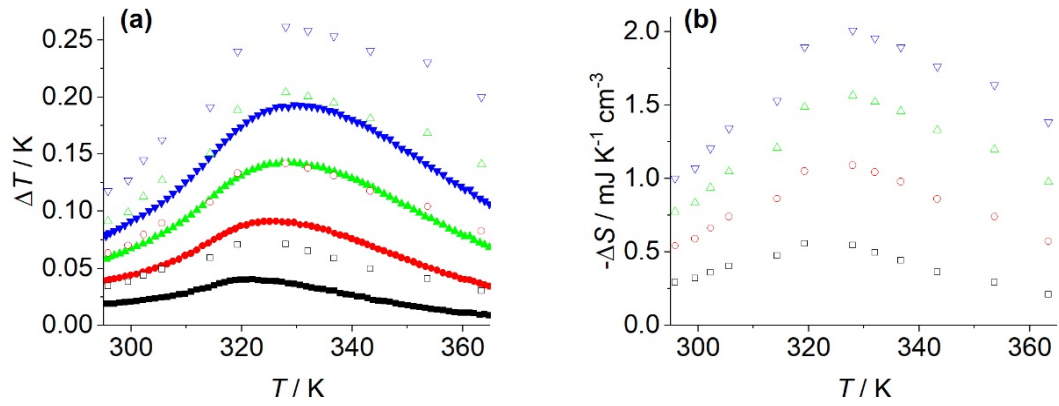

**FIG. 6**. For BCTH-6/15, direct (filled symbols) and indirect (empty symbols) ECE data, namely $\Delta T$ (a) and entropy change $-\Delta S$ (b), as a function of selected electric field changes: 4 (black), 8 (red), 12 (green), and 16 kV cm$^{-1}$ (blue).

Direct ECE measurements are complemented by the indirect ECE estimations that we calculate by applying the following equations to the experimental $P - E$ data (upper branches of the loops with $E > 0$), i.e.,

$$\Delta S = \int_0^E \left(\frac{\partial P}{\partial T}\right)_{E'} dE' \qquad (3)$$

$$\Delta T = -\frac{1}{\rho}\int_0^E \frac{T}{c_p}\left(\frac{\partial P}{\partial T}\right)_{E'} dE' \qquad (4)$$

for any field change $\Delta E = E - 0$, where $c_p$ is the heat capacity and $\rho$ the density. As widely accepted for practical $\Delta T$ calculations, $T$ in Eq. 4 is assumed field independent, and so is $c_p$ that is also typically taken temperature independent. Therefore, the $\Delta T$ calculation simplifies to

$$\Delta T \approx -\frac{T}{\rho c_p}\int_0^E \left(\frac{\partial P}{\partial T}\right)_{E'} dE' = -\frac{T\Delta S}{\rho c_p} \qquad (5)$$

for any field change $\Delta E = E - 0$. For $\rho$ = 6.3 g cm$^{-3}$, as deduced from the XRD refinements, and $c_p$ = 0.40 J K$^{-1}$ g$^{-1}$, as measured for the related compound BaTi$_{0.89}$Hf$_{0.11}$O$_3$,[16] the indirect estimations obtained for BCTH-6/15 (Fig. 6) and the other measured compositions (Figs. S6 and S7) predict the position of the maximum ECE. This can be seen by comparing direct $\Delta T$ data with indirect $\Delta S$ estimates (Eq. 3), both being characterized by an anomaly centered at a similar temperature. However, this agreement is not quantitatively satisfactory, as clearly manifested by the comparison between direct $\Delta T$ data and indirect $\Delta T$ estimates, based on Eq. 5. There are strong quantitative discrepancies that are also composition dependent. For instance, the maximum $\Delta T$ of BCTH-6/15 reaches 0.20 and 0.26 K for direct and indirect measurements, respectively (Fig. 6 for $E$ = 16 kV cm$^{-1}$), while, e.g., BCTH-6/12 behaves oppositely since the maximum $\Delta T$ = 0.23 and 0.16 K for direct and indirect measurements, respectively (Fig. S6 for $E$ = 14 kV cm$^{-1}$). Such non-systematic disagreement must be ascribed to the rough assumptions underlying Eq. 5.[22] Besides and not less importantly, the Maxwell's relations rely on thermodynamic equilibrium, which might not be reached especially for those compositions showing relaxor behavior, therefore making questionable their applicability.

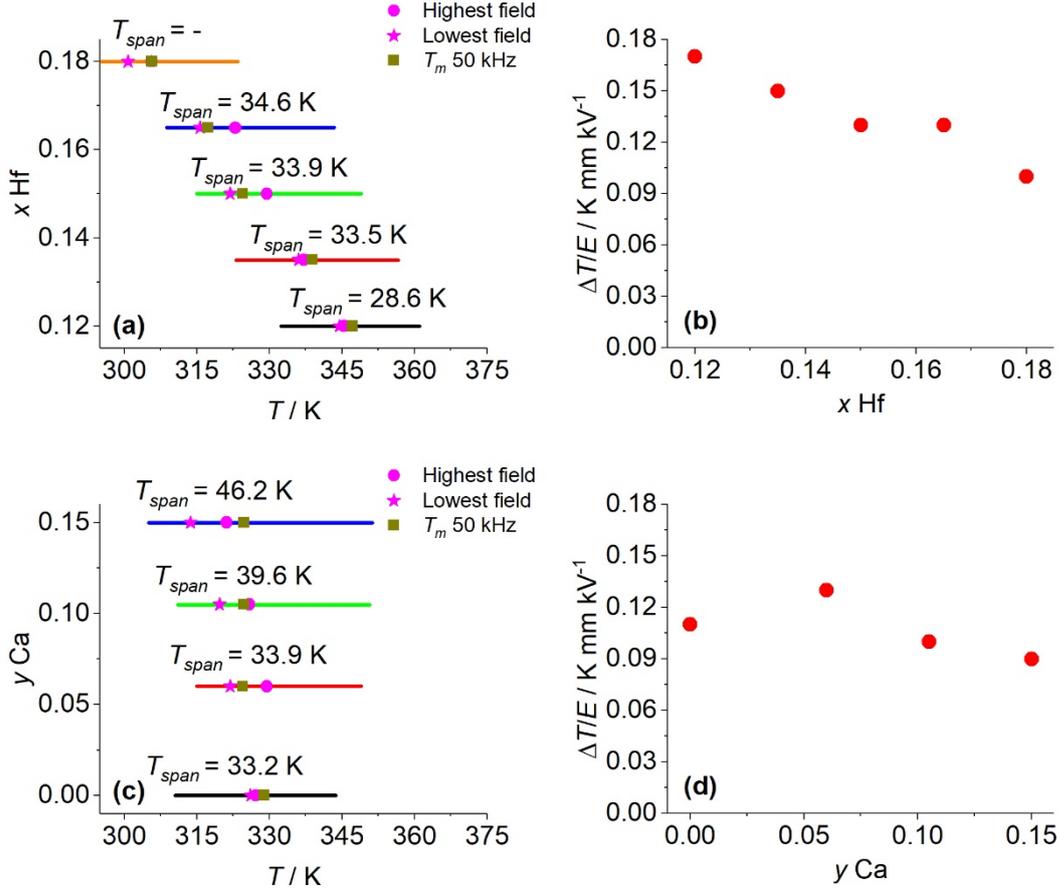

**FIG. 7**. Temperature span $T_{span}$ within which the ECE maintains 80% of its maximum value, and responsivity $\Delta T/E$ for BCTH-6/$x$ (a and b, respectively) and BCTH-$y$/15 (c and d, respectively). Shown also are the temperatures of the $\Delta T$ maxima for the highest and lowest applied electric fields, as well as the transition temperature $T_m$ determined from the dielectric permittivity data, at 50 kHz.

Finally, we look at the temperature span, $T_{span}$, within which the ECE maintains 80% of its maximum value, and the ECE responsivity, $\Delta T/E$, for any given composition (Fig. 7), as obtained from the direct ECE data. Note that Hf/Ca substitutions broaden $T_{span}$, in agreement with the behavior observed for the $\gamma$ exponent. Furthermore, the Hf/Ca substitutions also tend to increase the temperature difference between the $\Delta T$ maxima for the lowest (3/5 kV cm$^{-1}$) and highest (14/19 kV cm$^{-1}$) applied electric fields, respectively. The ECE responsivity tends to decrease with increasing Hf/Ca substitutions. This behavior should be attributed to the broadening of the transition, leading to a broader $T_{span}$ and an ECE overall weaker although closer to room temperature.

## IV. CONCLUSIONS

We employ the conventional solid-state reaction method to synthesize several compositions of $Ba_{1-y}Ca_yTi_{1-x}Hf_xO_3$ ($0.12 < x < 0.18$, $0 < y < 0.15$). We systematically investigate their dielectric and electrocaloric properties as a function of the Hf/Ca content. We find that the ferroelectric-paraelectric phase transition moves toward room temperature and broadens as the Hf or Ca content increases, the averaged Curie temperature reaching a value as low as $T_m$ = 306.2 K for $Ba_{0.94}Ca_{0.06}Ti_{0.82}Hf_{0.18}O_3$ (i.e., $x = 0.18$, $y = 0.06$). Two interferroelectric phase transitions are clearly identified for $Ba_{0.94}Ca_{0.06}Ti_{0.88}Hf_{0.12}O_3$ (i.e., $x = 0.12$, $y = 0.06$) and tend to converge, together with the ferroelectric-paraelectric phase transition, at ca. 335 K.

The electrocaloric effect is measured for all compositions with the use of direct and indirect methods. The electrocaloric temperature change reaches values of ~ 0.2 K for an applied field change of 16 kV cm$^{-1}$. While the addition of Hf/Ca weakens the ECE responsivity, it is successful in tuning the effect closer to room temperature and incrementing its temperature span. The direct electrocaloric measurements prove to be extremely valuable since the indirect method fails to determine reliably the strength of the electrocaloric effect.

## SUPPLEMENTARY MATERIAL

See the Supplementary Material for XRD patterns for BCTH-6/$x$ and BCTH-$y$/15 (Fig. S1); dielectric permittivity data for BCTH-6/$x$ (Fig. S2) and BCTH-$y$/15 (Fig. S3); polarization-electric field hysteresis loops for BCTH-6/$x$ (Fig. S4) and BCTH-$y$/15 (Fig. S5); and comparisons between direct and indirect electrocaloric data for BCTH-6/$x$ (Fig. S6) and BCTH-$y$/15 (Fig. S7).

## ACKNOWLEDGEMENTS

This work was supported by MICINN (project No. PID2021-124734OB-C21) and DGA (E11-23R, E12-23R). S.L. acknowledges funding from the European Union's Horizon 2020 research and innovation program under the Marie Skłodowska-Curie grant agreement No. 101029019. D.G. acknowledges financial support from the Gobierno de

Aragón through a doctoral fellowship. We thank the Servicio General de Apoyo a la Investigación from the Universidad de Zaragoza.

**AUTHOR DECLARATIONS**

**Conflict of Interest**

The authors have no conflicts to disclose.

**Author Contributions**

**David Gracia**: Data curation (lead); Investigation (equal); Methodology (equal); Writing – original draft (lead); Writing – review & editing (equal). **Sara Lafuerza**: Data curation (supporting); Investigation (equal); Funding acquisition (supporting); Methodology (equal); Writing – original draft (supporting); Writing – review & editing (equal). **Javier Blasco**: Data curation (supporting); Investigation (equal); Funding acquisition (lead); Supervision (equal); Methodology (equal); Writing – original draft (supporting); Writing – review & editing (equal). **Marco Evangelisti**: Data curation (supporting); Investigation (equal); Funding acquisition (supporting); Supervision (equal); Methodology (equal); Writing – original draft (supporting); Writing – review & editing (equal).

**DATA AVAILABILITY**

The data that support the findings of this study are available from the authors upon reasonable request.

**APPENDIX**

**Homemade quasi-adiabatic calorimeter for direct electrocaloric measurements**

A homemade quasi-adiabatic calorimeter is used for the ECE direct determination from 4.2 to 370 K, although only temperatures above room temperature are explored for this work. The setup primarily consists of an outer vacuum chamber (OVC) and an inner chamber. The OVC is sealed by indium junctions and its pressure stabilizes to around $10^{-6}$ mbar, or below, during the measurements. The inner chamber is delimited by a thermal shield provided with a resistive heater and a T-type thermocouple, to regulate the temperature of the measurements with the help of an Oxford Instruments MercuryiTC temperature controller. The thermal shield is mechanically anchored to an oxygen-free copper block, whose temperature is measured by a PT100 sensor and a Keithley 2000 multimeter. Inside the inner chamber, the disc-shaped sample is in "contactless" mode, that is, the sample is held in place by the electrical wiring only (Fig. 8).

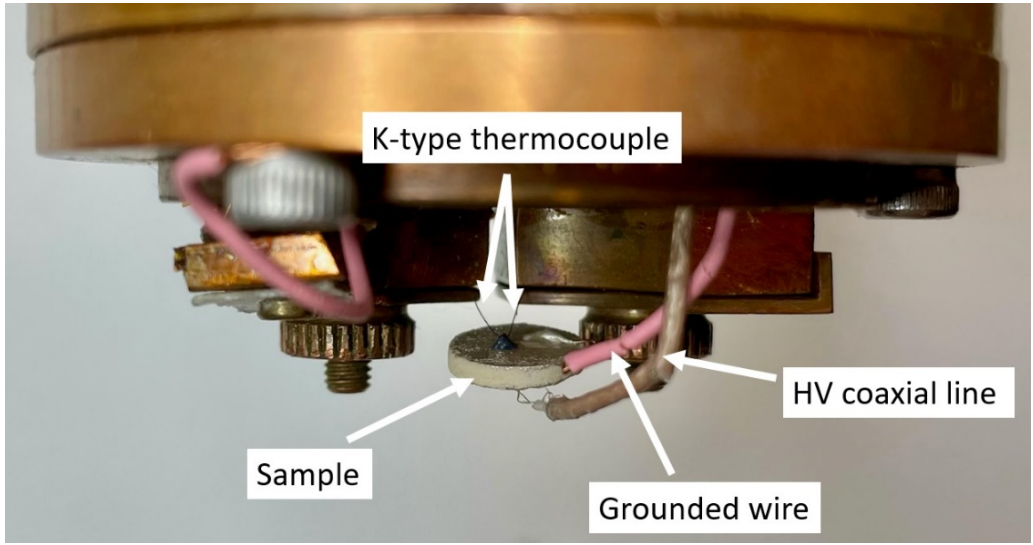

**FIG. 8**. Inside of the inner chamber of the homemade quasi-adiabatic calorimeter.

On the disc-shaped sample, the voltage is applied to one side with a high-voltage (HV) coaxial line, while a grounded wire is contacted to the opposite side. The electrical contacts are made with silver epoxy, Epo-Tek H20E. We make use of a homemade voltage source, whose output is 0 – 2 kV. A K-type thermocouple measures the temperature difference between the grounded side of the sample and the copper block, with the help of a Keithley 2182A nanovoltmeter. This thermocouple is made of extremely thin leads (diameter of 0.003 inch, from OMEGA Engineering) to minimize thermal losses, and bonded to the sample by a thermally conducting adhesive that prevents electrical contact, Multicomponent MC002964. Care is taken to avoid unneeded junctions that can produce parasitic voltages in the thermocouple. All electronics are controlled with a computer and a LabVIEW based software, so that the whole process is operated automatically.

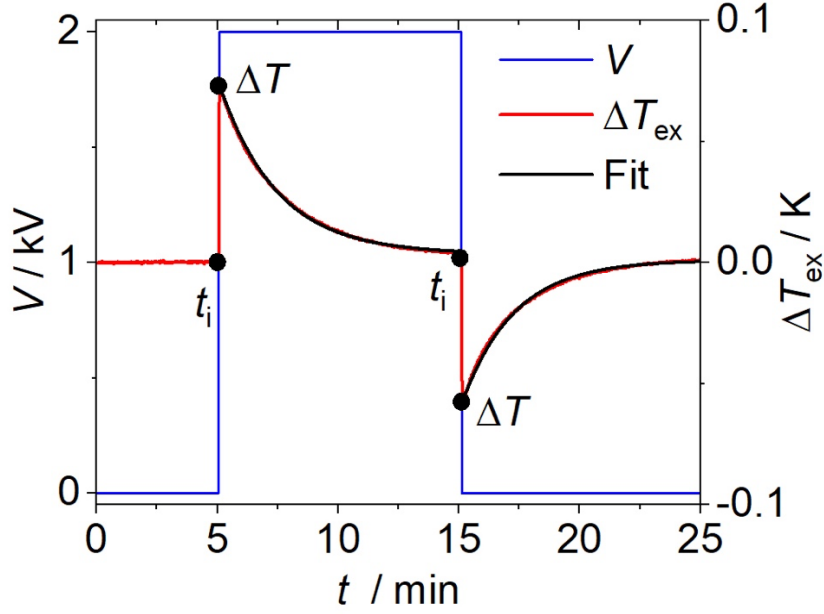

**FIG. 9**. Representative direct electrocaloric measurement as a function of time for a polarization/depolarization sequence for BCTH-15/15, at $V = 2$ kV and $T_{bath} = 294$ K. The fit is based on an exponential decay (Eq. A1).

In relaxation experiments, we monitor the temperature difference, $\Delta T_{ex}(t)$, between the sample and the copper block at any time $t$. Let us start the measurement at thermal equilibrium (Fig. 9). The ECE kicks in almost instantly, therefore quasi-adiabatically, following the polarization/depolarization, i.e., the on/off switching of the electric field, $E = V/d$. After each jump, the temperature of the sample relaxes to the temperature, $T_{bath}$, of the thermal bath, i.e., the copper block. The relaxation can be well described by an exponential decay,

$$\Delta T_{ex} = \Delta T\, e^{-(t-t_i)/\tau} \qquad (A1)$$

where $\Delta T$ is the equivalent temperature change under ideal adiabatic conditions, $t_i$ indicates the instant when $E$ is applied (or withdrawn), and $\tau$ is the characteristic relaxation time. Depending on sample, temperature, and applied field, we obtain values of $\tau$ up to 4 minutes (e.g., from Fig. 9, $\tau = 2.3$ and 2.1 min for the polarization and depolarization, resp., for $V = 2$ kV and $T_{bath} = 294$ K), that is, the measurement of an individual sequence for full polarization/depolarization at thermal equilibrium takes more than 20 minutes to complete. Since this time scale makes unfeasible the collection of such data for several samples under many different experimental conditions (temperatures and applied fields), we adopt the following measurement protocol. We stabilize the temperature of the copper block in 1 K-steps. At every temperature step, we apply pulses of voltages $V$ = 0.5, 1.0, 1.5 and 2.0 kV, respectively, being each pulse 100 sec in length and repeated twice, while monitoring $\Delta T_{ex}(t)$ continuously. A representative multiple-pulse sequence measurement is depicted in Fig. 10. For such relatively short times scales (100 sec is significantly shorter than $\tau$'s), the system does not reach the thermal equilibrium between pulses and the exponential fitting function (Eq. A1) must be modified to

$$\Delta T_{ex} = \Delta T_\infty + [\Delta T - (\Delta T_\infty - \Delta T_i)]e^{-(t-t_i)/\tau} \qquad (A2)$$

where $\Delta T_i$ is experimentally known and corresponds to $\Delta T_{ex}$ just before the jump, while the adiabatic temperature $\Delta T$, the characteristic relaxation time $\tau$, and $\Delta T_\infty = \Delta T_{ex}(t \to t_\infty)$ are fitting parameters, for any sample temperature $T = T_{bath} + \Delta T_i$.

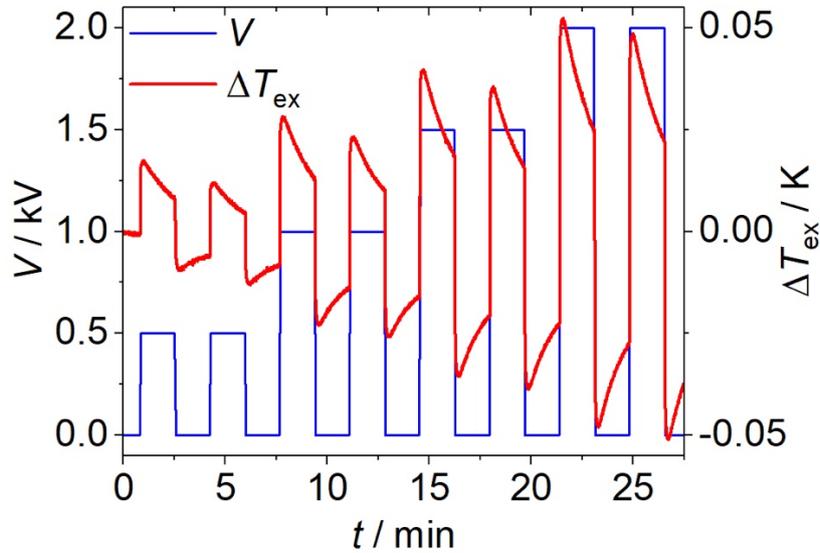

**FIG. 10**. Representative direct electrocaloric measurement as a function of time for a voltage multiple-pulse sequence for BCTH-6/15, at $T_{bath}$ = 300 K.

The mass of the addenda (adhesives, wires) is less than 5 % of the sample mass. Accordingly, the addenda heat capacity is negligible with respect to the sample contribution and corrections to $\Delta T$ can safely be disregarded.

Supplementary Material

# The electrocaloric effect of lead-free Ba$_{1-y}$Ca$_y$Ti$_{1-x}$Hf$_x$O$_3$ from direct and indirect measurements

David Gracia, Sara Lafuerza, Javier Blasco, and Marco Evangelisti

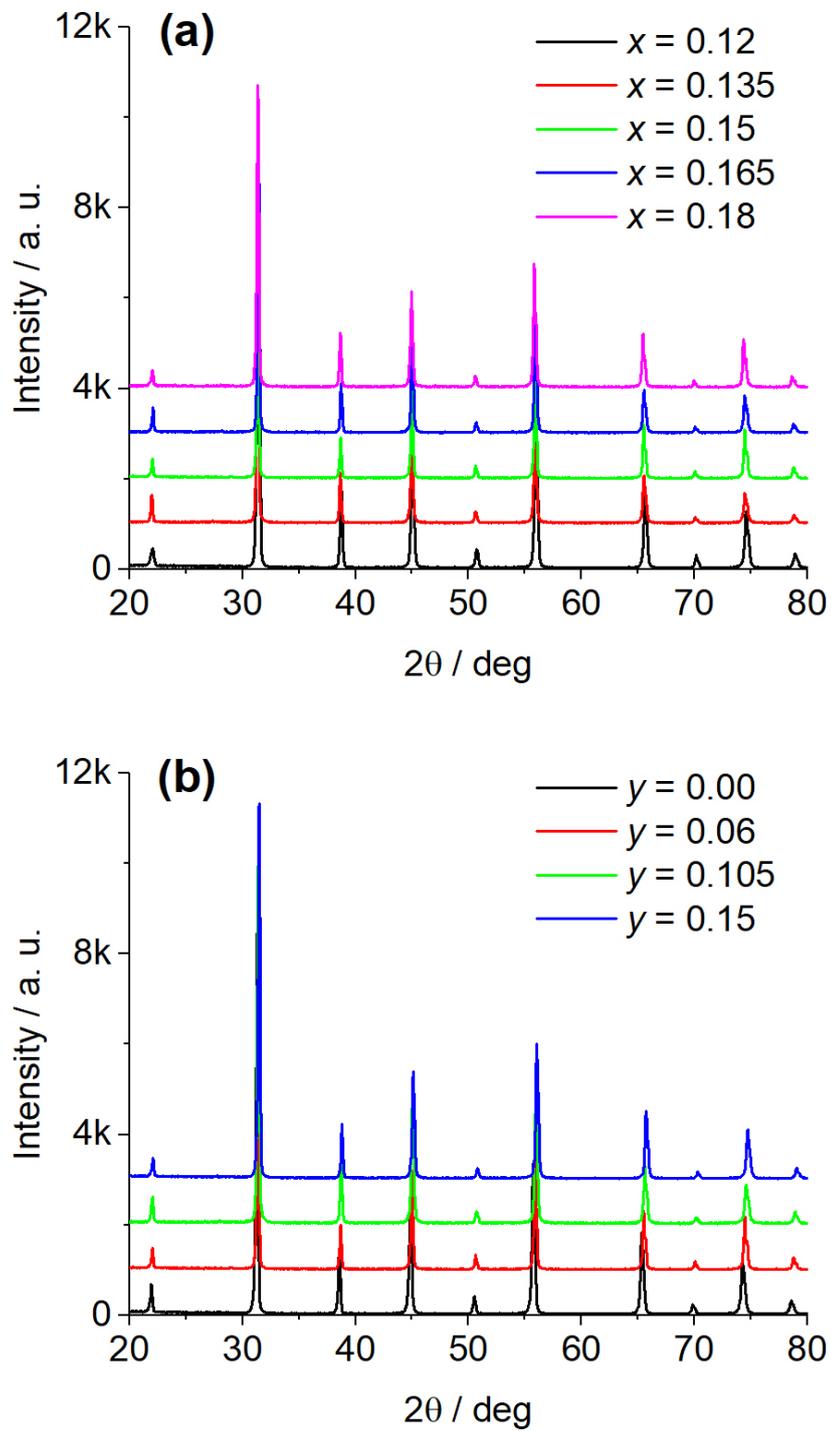

**FIG. S1.** XRD patterns for BCTH-6/*x* (a) and BCTH-*y*/15 (b).

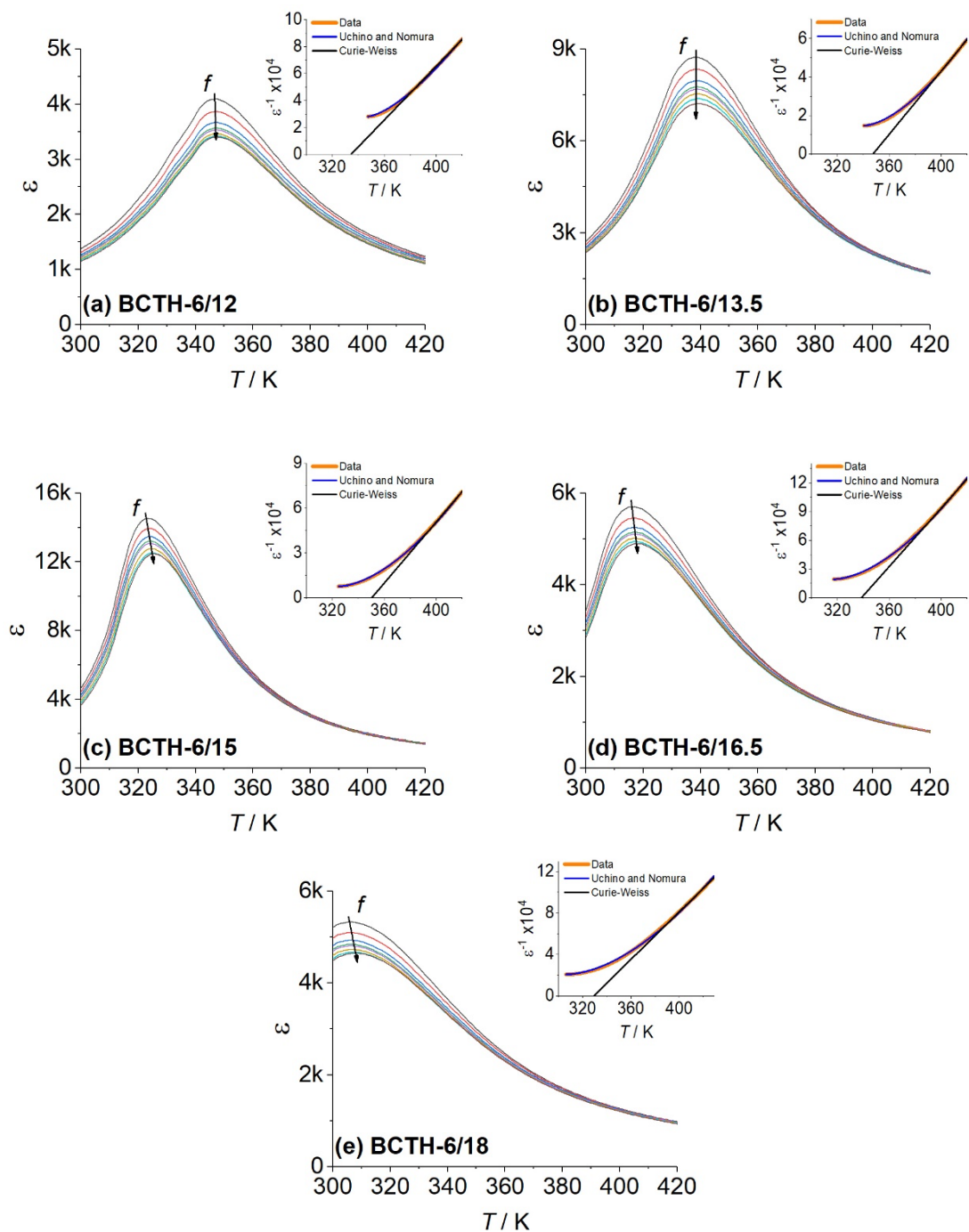

**FIG. S2.** Temperature dependence of the real component of the dielectric permittivity $\varepsilon$ for each investigated BCTH-6/*x* composition, as labelled, measured for frequencies *f* ranging from 100 Hz to 4 MHz, together with the Curie-Weiss and Uchino and Nomura fits for $f$ = 50 kHz (insets).

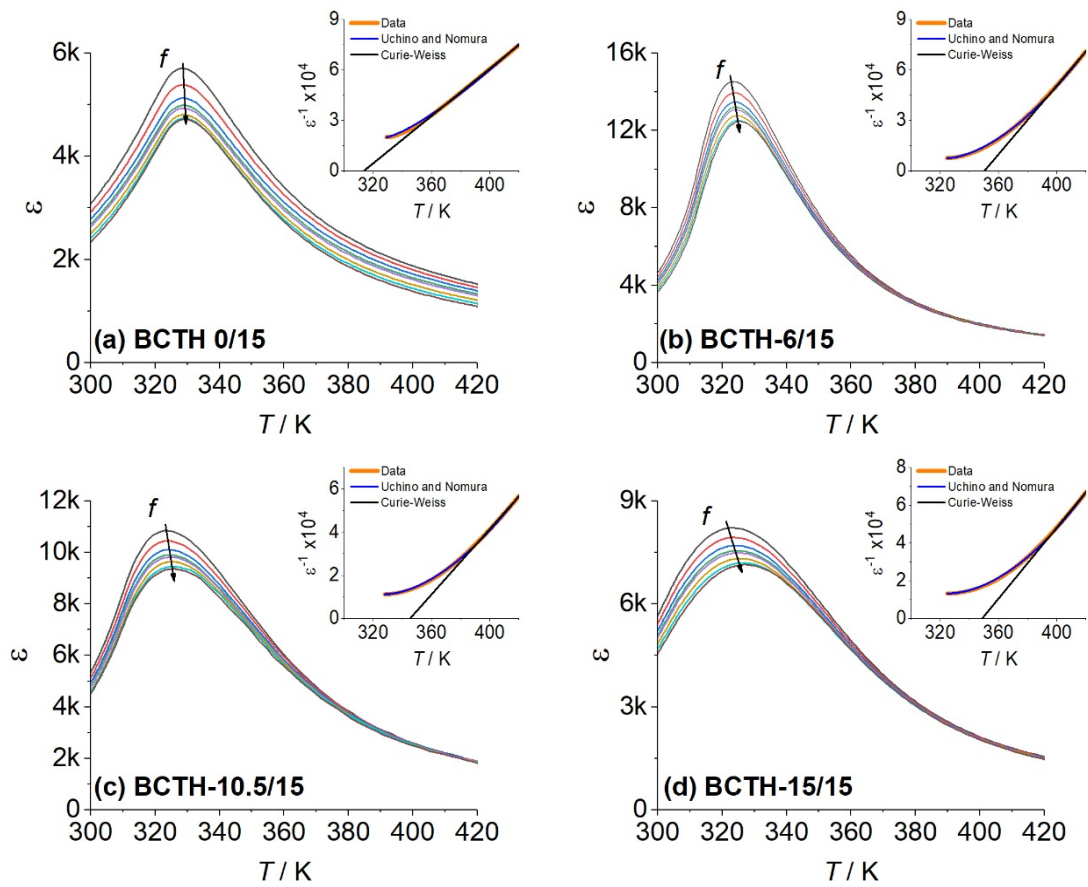

**FIG. S3.** Temperature dependence of the real component of the dielectric permittivity $\varepsilon$ for each investigated BCTH-*y*/15 composition, as labelled, measured for frequencies *f* ranging from 100 Hz to 4 MHz, together with the Curie-Weiss and Uchino and Nomura fits for $f$ = 50 kHz (insets).

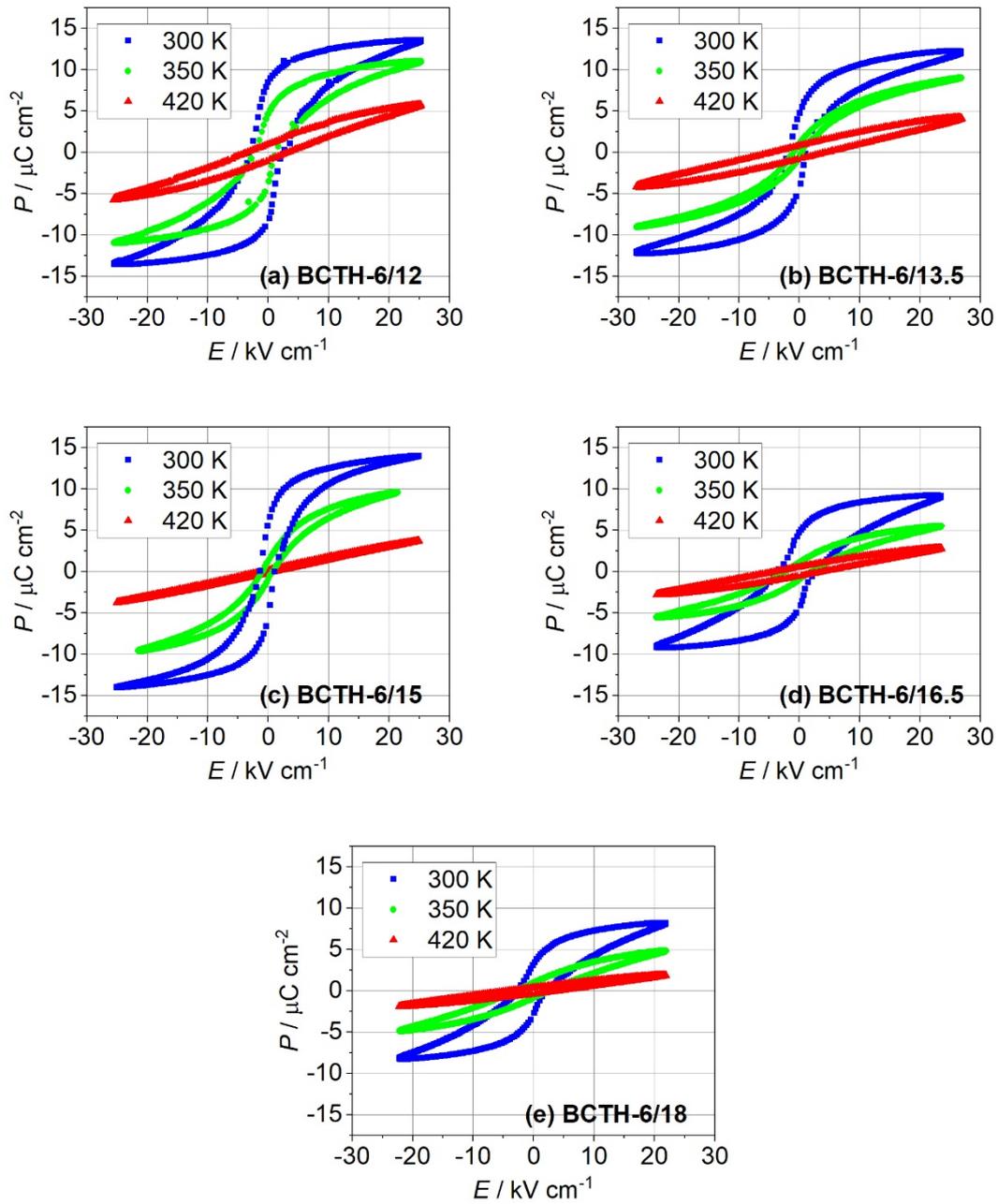

**FIG. S4.** Polarization-electric field hysteresis loop for each investigated BCTH-6/*x* composition, at three selected temperatures, as labelled.

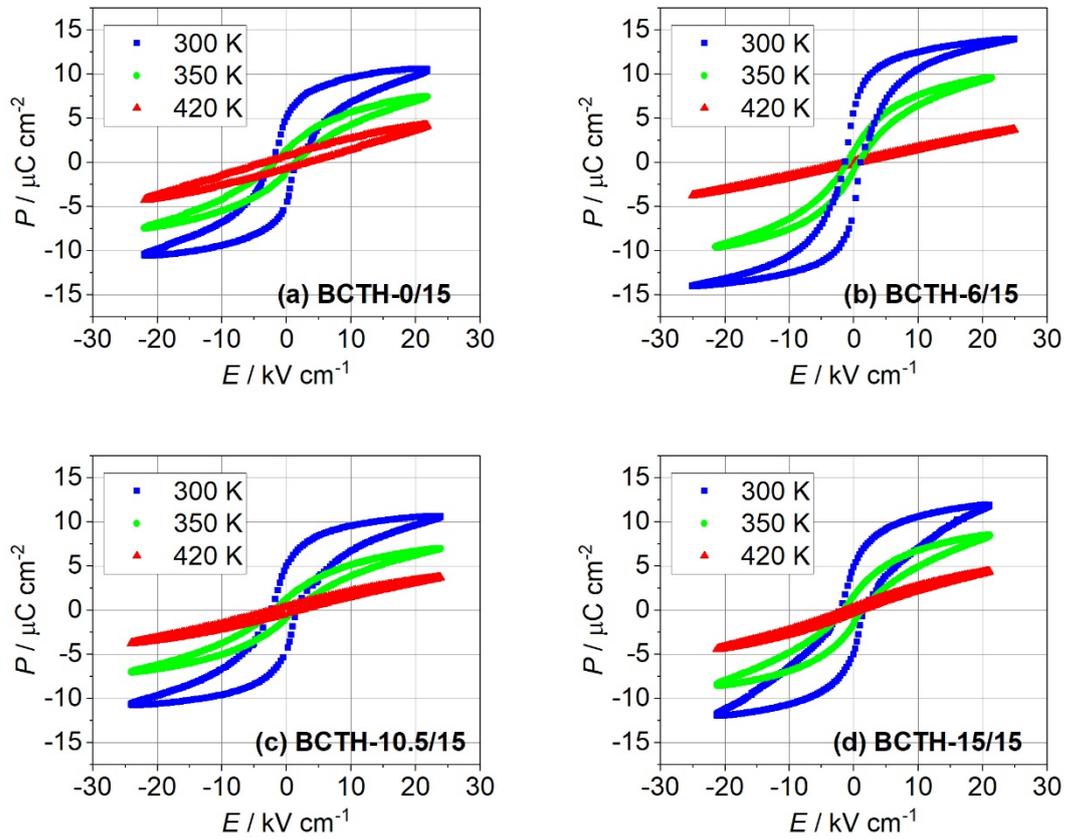

**FIG. S5.** Polarization-electric field hysteresis loop for each investigated BCTH-*y*/15 composition, at three selected temperatures, as labelled.

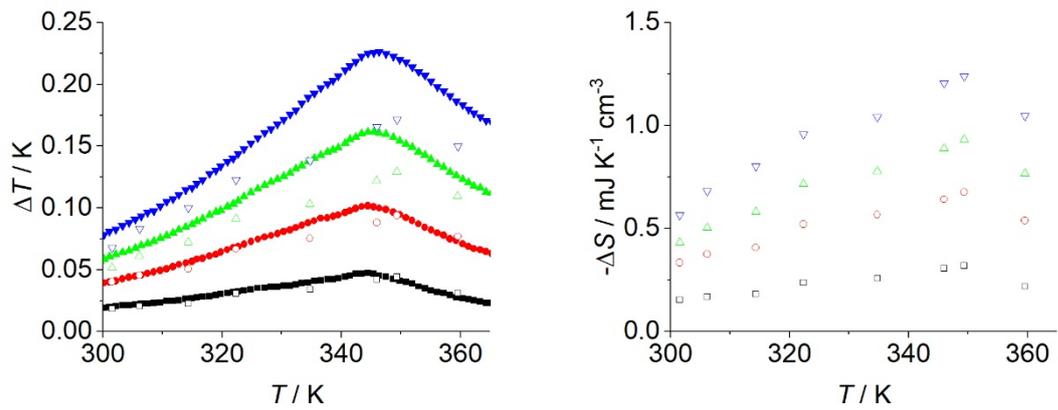

**BCTH-6/12.** Electric fields: black: 3, red: 7, green: 10, blue: 14, in kV cm$^{-1}$.

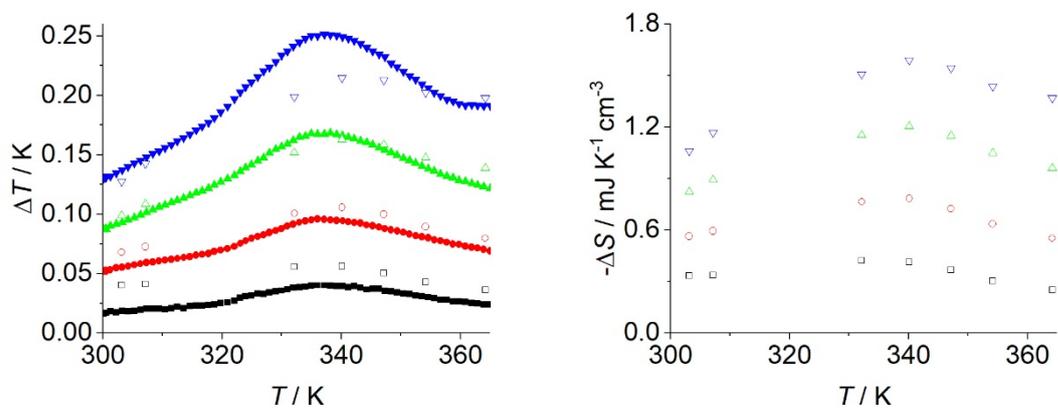

**BCTH-6/13.5.** Electric fields: black: 5, red: 9, green: 14, blue: 19, in kV cm$^{-1}$.

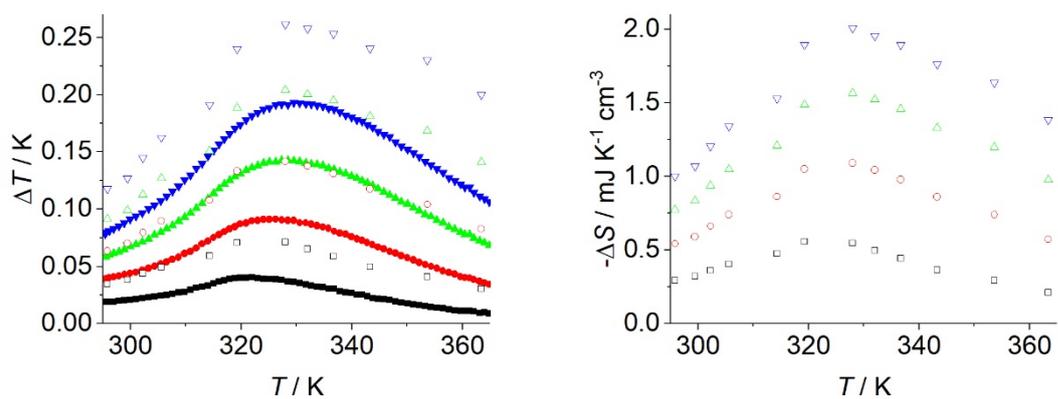

**BCTH-6/15.** Electric fields: black: 4, red: 8, green: 12, blue: 16, in kV cm$^{-1}$.

( *continue below* )

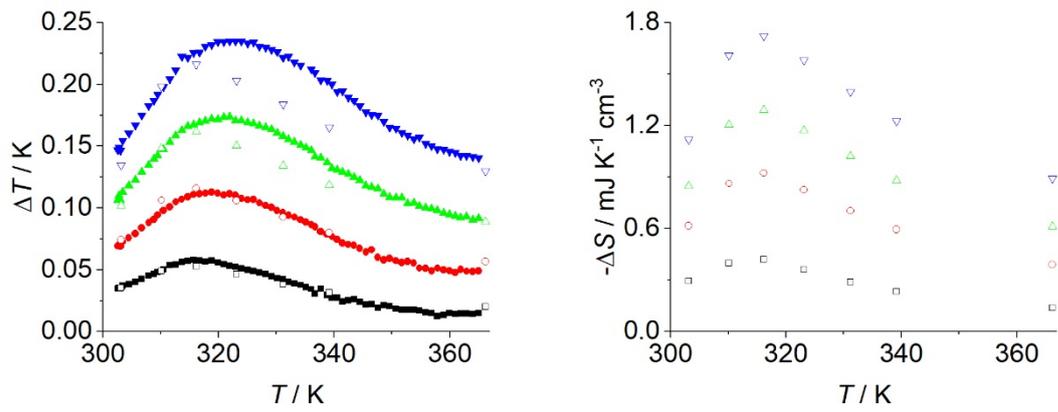

**BCTH-6/16.5.** Electric fields: black: 4, red: 9, green: 13, blue: 18, in kV cm$^{-1}$.

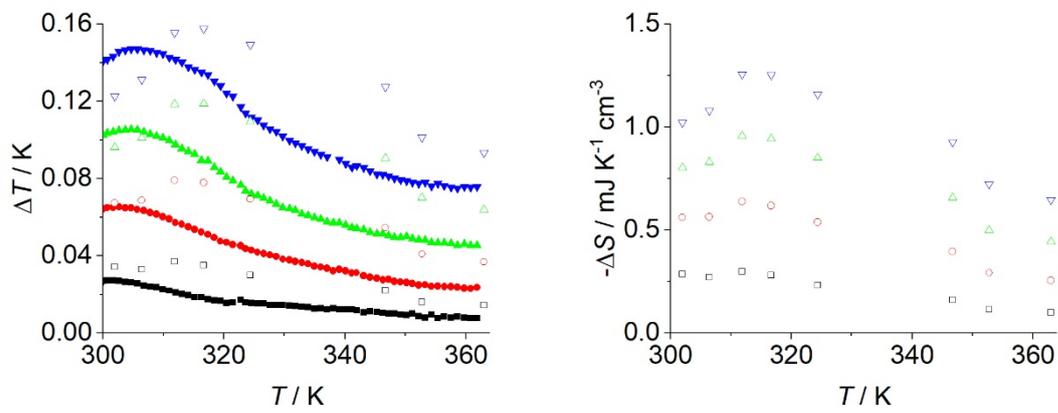

**BCTH-6/18.** Electric fields: black: 4, red: 8, green: 12, blue: 16, in kV cm$^{-1}$.

**FIG. S6.** Direct (filled symbols) and indirect (empty symbols) electrocaloric data, namely adiabatic temperature change $\Delta T$ and entropy change $-\Delta S$, for each investigated BCTH-6/$x$ composition, as a function of several electric field values, as indicated.

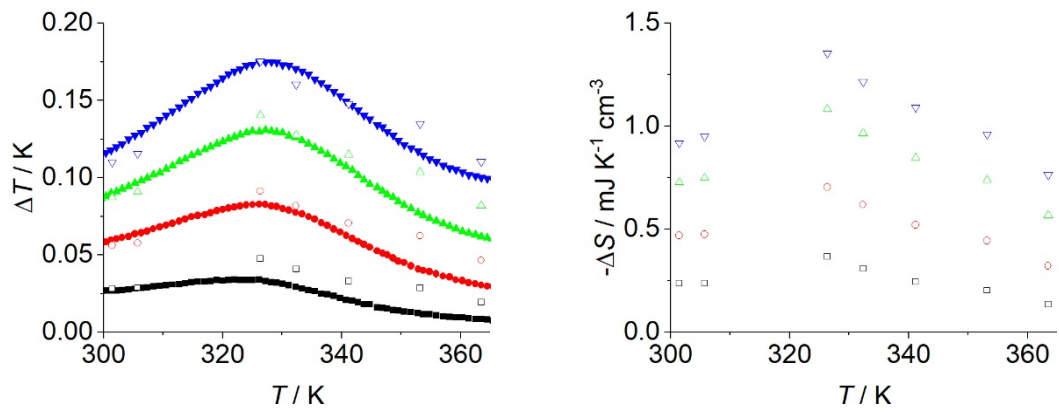

**BCTH-0/15.** Electric fields: black: 4, red: 8, green: 13, blue: 17, in kV cm$^{-1}$.

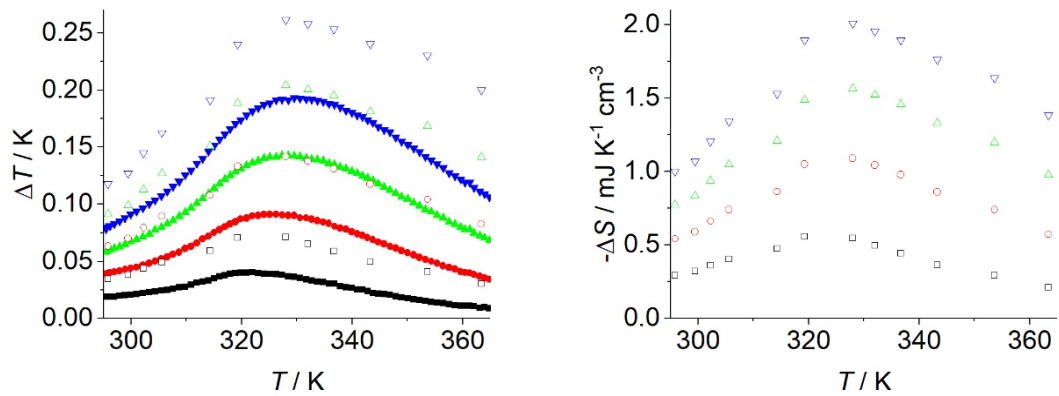

**BCTH-6/15.** Electric fields: black: 4, red: 8, green: 12, blue: 16, in kV cm$^{-1}$.

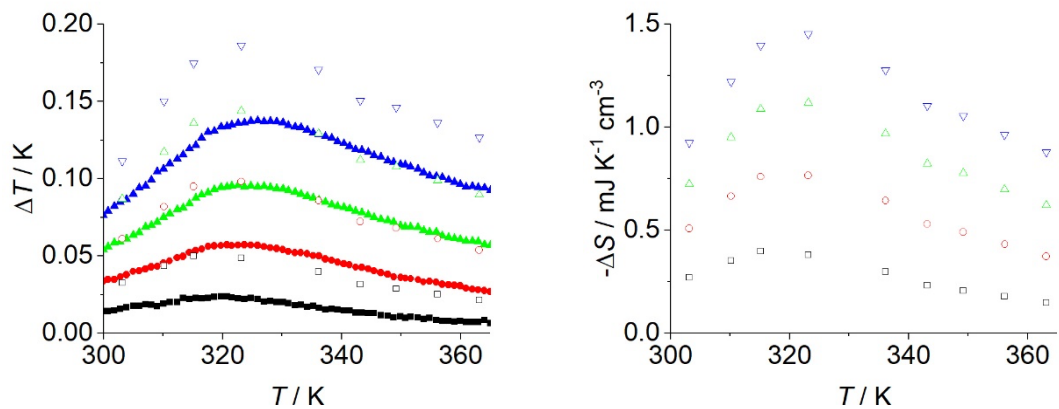

**BCTH-10.5/15.** Electric fields: black: 4, red: 8, green: 12, blue: 16, in kV cm$^{-1}$.

( *continue below* )

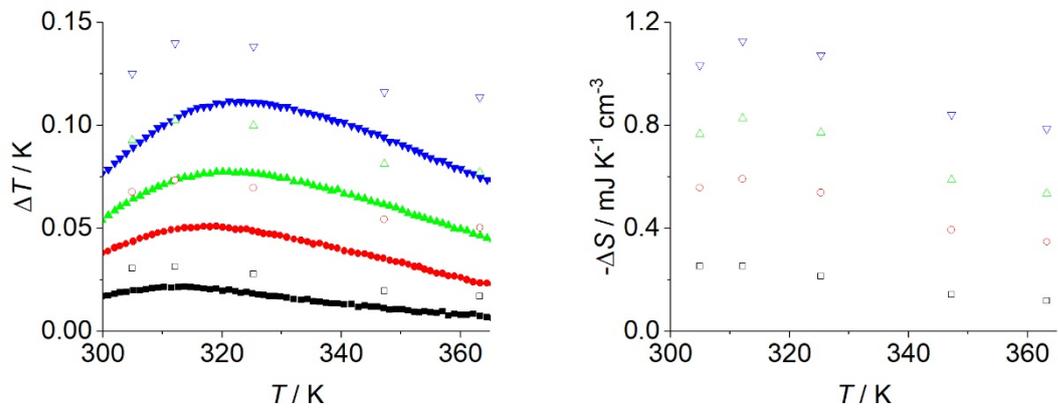

**BCTH-15/15.** Electric fields: black: 3, red: 7, green: 10, blue: 14, in kV cm$^{-1}$.

**FIG. S7.** Direct (filled symbols) and indirect (empty symbols) electrocaloric data, namely adiabatic temperature change $\Delta T$ and entropy change $-\Delta S$, for each investigated BCTH-*y*/15 composition, as a function of several electric field values, as indicated.